\documentclass[sigconf]{acmart}
\usepackage{algorithm}
\usepackage{algpseudocode}
\newcommand{\ie}{\textit{i.e.},}
\usepackage{amsmath,amsfonts}
\usepackage{graphicx}
\usepackage{textcomp}
\usepackage{xcolor}
\usepackage{subfigure}
\usepackage{multirow}
\def\BibTeX{{\rm B\kern-.05em{\sc i\kern-.025em b}\kern-.08em
    T\kern-.1667em\lower.7ex\hbox{E}\kern-.125emX}}

\usepackage{tikz}
\usepackage{verbatim}
\usetikzlibrary{calc,trees,positioning,arrows,chains,shapes.geometric,%
    decorations.pathreplacing,decorations.pathmorphing,shapes,%
    matrix,shapes.symbols}
\usepackage[linewidth=1pt]{mdframed}
\tikzset{
    >=stealth',
  punktchain/.style={
    rectangle, 
    rounded corners, 
    draw=black, very thick,
    text width=10em, 
    minimum height=3em, 
    text centered, 
    on chain},
  line/.style={draw, thick, <-},
  element/.style={
    tape,
    top color=white,
    bottom color=blue!50!black!60!,
    minimum width=8em,
    draw=blue!40!black!90, very thick,
    text width=10em, 
    minimum height=3.5em, 
    text centered, 
    on chain},
  every join/.style={->, thick,shorten >=1pt},
  decoration={brace},
  tuborg/.style={decorate},
  tubnode/.style={midway, right=2pt},
}

\begin{document}

\title{Improving Code Example Recommendations on Informal Documentation Using BERT and Query-Aware LSH: A Comparative Study
}

\author{Sajjad Rahmani}
\email{sajjad.rahmani.1@ens.etsmtl.ca}
\affiliation{%
  \institution{École de Technologie Supérieure}
  \city{Montreal}
  \state{Quebec}
  \country{Canada}
}
\author{AmirHossein Naghshzan}
\email{amirhossein.naghshzan.1@ens.etsmtl.ca}
\affiliation{%
  \institution{École de Technologie Supérieure}
  \city{Montreal}
  \state{Quebec}
  \country{Canada}
}
\author{Latifa Guerrouj}
\email{latifa.guerrouj@etsmtl.ca}
\affiliation{%
  \institution{École de Technologie Supérieure}
  \city{Montreal}
  \state{Quebec}
  \country{Canada}
}

\begin{abstract}
Our research investigates the recommendation of code examples to aid software developers, a practice that saves developers significant time by providing ready-to-use code snippets. The focus of our study is Stack Overflow, a commonly used resource for coding discussions and solutions, particularly in the context of the Java programming language.

We applied BERT, a powerful Large Language Model (LLM) that enables us to transform code examples into numerical vectors by extracting their semantic information. Once these numerical representations are prepared, we identify Approximate Nearest Neighbors (ANN) using Locality-Sensitive Hashing (LSH). Our research employed two variants of LSH: Random Hyperplane-based LSH and Query-Aware LSH. We rigorously compared these two approaches across four parameters: HitRate, Mean Reciprocal Rank (MRR), Average Execution Time, and Relevance.

Our study revealed that the Query-Aware (QA) approach showed superior performance over the Random Hyperplane-based (RH) method. Specifically, it exhibited a notable improvement of 20\% to 35\% in HitRate for query pairs compared to the RH approach. Furthermore, the QA approach proved significantly more time-efficient, with its speed in creating hashing tables and assigning data samples to buckets being at least four times faster. It can return code examples within milliseconds, whereas the RH approach typically requires several seconds to recommend code examples. Due to the superior performance of the QA approach, we tested it against PostFinder and FaCoY, the state-of-the-art baselines. Our QA method showed comparable efficiency proving its potential for effective code recommendation.

\end{abstract}


\maketitle

\section{Introduction}
Recommendation systems have become ubiquitous in various fields to enhance task efficiency and quality. Zhou {\em et al.}~\cite{zhou2019lancer} note that software developers frequently write similar code examples multiple times due to the need to implement comparable functionalities in different projects. Therefore, during the software development process, a recommendation system can assist programmers in completing their tasks quickly and effectively by presenting them with the most pertinent and high-quality examples written by other programmers  \cite{di2021development}. Open-source projects and informal documentation are the two main sources of information that developers rely on to perform programming tasks. For instance, GitHub provides open-source projects that offer code examples for various tasks and code resources for use.

Informal documentation, as opposed to Open-Source projects, comprises data sources that developers use to exchange information and ideas about various tasks. These sources may include incomplete code examples such as bug reports, emails, and Stack Overflow posts \cite{kim2018facoy}. However, code examples in informal documentation are typically trapped in natural language comments, which can make it challenging to extract relevant code entities and elements. 
Therefore, the objective of this paper is to propose a recommendation system that can suggest relevant code examples from Stack Overflow based on developers' needs and tasks. Stack Overflow was chosen as the data source for the recommendation system as it is one of the most popular resources among developers for addressing programming issues \cite{rubei2020postfinder}.\\
The research specifically targets Java code examples posted on Stack Overflow in response to questions, between 2008 and May 2022, and preserved in dump files. Once pre-processing is done, the code examples are translated into numerical vectors using the BERT model, followed by the execution of two LSH-based algorithms: Random Hyperplane-based LSH and Query-Aware LSH, to minimize the search space and detect code examples that are similar to a user's query.\\
Our decision to use both BERT and LSH was motivated by two key factors. Firstly, BERT has demonstrated its ability to extract semantic information from natural language texts in various applications, including cross-lingual translation \cite{devlin2018bert, wolf2020transformers}. We believe that this capability can also be leveraged to extract semantic information from code examples and identify similarities between them. Secondly, due to the large amount of data samples exceeding 60K, it can be a tedious task to locate relevant code examples. However, LSH-based approaches like Random Hyperplane-based and Query-Aware can be utilized to minimize the search space and facilitate the process of discovering relevant code examples.\\
Our focus in this paper is to evaluate the performance of our LSH-based algorithms by exploring three main research questions:

\textit{\textbf{RQ1:}} How do the \emph{Random Hyperplane-based LSH} and \emph{Query-Aware LSH} algorithms perform in recommending code examples, with regards to metrics such as \emph{HitRate, Mean Reciprocal Rank, Average execution time}, and \emph{Relevance}?

\textit{\textbf{RQ2:}} Do any significant differences exist in the \emph{Relevance} values between the \emph{Random Hyperplane-based LSH} and \emph{Query-Aware LSH} algorithms?

\textit{\textbf{RQ3:}} How does the proposed algorithm perform compared to the recent state-of-the-art methods, namely PostFinder and FaCoY?\\

\section{Related Work}
In this part, we have categorized the related work into different sections which are presented as subsections. The first section covers the proposed works on extracting code elements from informal documentation and identifying patterns of API methods that are frequently used together. The next category highlights research on recommending code examples from open-source projects. Finally, we discuss some research works that have utilized LSH in suggesting recommendation systems.

\subsection{Informal Documentation}
InfoZilla \cite{bettenburg2008extracting}, aims to extract various components from bug reports, including patches, Stack traces, source code, and enumerations. To extract code elements, this approach employs an island parser based on the concept of islands within the sea, which searches for islands based on identifiers. The term "island" in this context refers to classes, conditional statements, functions, and assignments that can be used to locate code elements in discussions. The island parser used by InfoZilla is based on the works of Moonen {\em et al.}~\cite{moonen2001generating} and Bacchelli {\em et al.}~\cite{bacchelli2011extracting}. 
Our approach differs from this research in that we do not utilize an island parser to identify code entities within discussions. Instead, we focus on extracting code examples from Stack Overflow posts that are enclosed by <code> tags. The purpose of this effort is to gather and suggest high-quality code examples according to the needs of developers.

ACE \cite{rigby2013discovering} utilizes the island parser concept along with naming conventions, such as Camel Case, to extract code elements that are embedded within informal documentation. This approach employs an island parser to locate Java code elements within Stack Overflow posts. ACE searches for qualified terms in the posts, including package names, variable declarations, and qualified variables, as well as class concepts like inheritance, constructors, and exceptions. 

Diamantopoulos {\em et al.}~\cite{diamantopoulos2015employing} have proposed an approach to extract code entities from Stack Overflow posts. This approach extracts three types of entities from Stack Overflow, \ie{} Java code examples, including Assignments (AM), Function calls (FC), and Class instantiations (CI). Our approach utilizes the attention mechanism \cite{niu2021review} implemented in the BERT model to identify similar code examples based on their semantic similarity rather than relying solely on code entities such as classes, variables, and functions.\\
Naghshzan {\em et al.}~\cite{Naghshzan-a,Naghshzan-b} presented a new approach to generating natural language summaries for Android API methods using Stack Overflow discussions. The approach was evaluated through a survey of 16 developers and found to be a useful complementary source of information for software development and maintenance tasks. The study contributes to the field of code summarization and highlights the potential of unofficial documentation in the process. Furthermore, in their latest research, they applied BERT for topic modeling and categorized the problems and potential solutions of Android APIs~\cite{Naghshzan-c}.

Abdalkareem {\em et al.}~\cite{abdalkareem2017code} have proposed a method for extracting code examples from Stack Overflow posts by filtering special tags. The technique involves encoding the bodies of Stack Overflow posts in HTML, which allows for the identification of code examples embedded within \textless code\textgreater tags through a process of filtering. We have applied this approach in our work as our dataset had the same structure as its and the code element extraction method was applicable in our work for filtering code examples that were trapped inside Stack Overflow posts. 

Kim {\em et al.}~\cite{kim2018facoy} presented FaCoY, which is a code-to-code recommendation system designed to identify code examples that are semantically similar to the query code. This work suffers from the drawback of producing a significant number of false positives, \ie{} returning a large number of irrelevant Stack Overflow posts as a result. 

The study by Rubei {\em et al.}~\cite{rubei2020postfinder} introduces PostFinder, a recommendation system plugin for Eclipse IDE that extracts contextual information from a developing project and Stack Overflow posts to provide recommendations for software developers. 

\subsection{Open-source projects}
Nguyen {\em et al.}~\cite{nguyen2019focus} proposed FOCUS (API FunctiOn Calls and USage patterns) as a method to suggest a group of methods that are frequently utilized together by developers while working with particular APIs in their development projects. The suggested methods are accompanied by usage patterns that serve as a reference for developers to complete their coding tasks. This approach recommends API methods based on the collaborative filtering concept. Using FOCUS, developers receive recommendations for relevant API invocations and code examples as useful references. 

Wang {\em et al.}~\cite{wang2013mining} introduced UP-Miner (Usage-pattern Miner), which mines commonly used API methods from source code by applying BI-Directional Extension(BIDE) algorithm \cite{wang2004bide}.

Gu {\em et al.}~\cite{gu2016deep} developed DeepAPI, a deep learning-based approach that generates API usage sequences by applying RNN Encoder-Decoder\cite{medsker2001recurrent}. DeepAPI has relied on GitHub and JavaDoc to provide recommendations for code examples and overlooked valuable sources of informal documentation such as Stack Overflow. Our method, on the other hand, takes advantage of Stack Overflow by utilizing code examples tagged as answers to questions.

Raychev {\em et al.}~\cite{raychev2014code} proposed an approach for code completion that utilizes neural networks. The primary concept of their approach is to employ a natural language processing model to forecast probabilities of sentences while searching for sequences of method invocations to complete the code. 



Zhou {\em et al.}~\cite{zhou2019lancer} developed Lancer, a code-to-code recommendation system that utilizes the BERT model to address the Out Of Vocabulary (OOV) problem in code recommendation. Additionally, the approach employs the PageRank algorithm\cite{rogers2002google} to discover relevant libraries that are used together. In contrast to Lancer, which utilizes open-source projects, our approach concentrates on recommending code examples from Stack Overflow. Additionally, we employ a hybrid approach that involves the use of both BERT and LSH algorithms, as opposed to PageRank and BERT. Although our methodology differs from theirs, we employ the same evaluation metrics, specifically the HitRate parameter, to assess the effectiveness of our recommended code examples.

\subsection{Locality Sensitive Hashing for Recommendation Systems}
Ding {\em et al.}~\cite{ding2016kam1n0} utilized a combination of graph search and Locality-Sensitive Hashing (LSH) to locate similar instances of assembly code. Specifically, their approach involved using a hybrid strategy that relied on assembly code sources, especially when the main source code was not available. This technique, referred to as Adaptive LSH (ALSH), employed a tree structure to conduct the search. Upon receiving a query (q), the algorithm located the leaf node in various prefix trees. After a sufficient number of points had been identified, the subtrees were divided, and the search moved up a level, eliminating the least similar samples to decrease the search space.

Silavong {\em et al.}~\cite{silavong2022senatus} introduced a code-to-code recommendation system that employs Locality Sensitive Hashing. The research utilizes ANTLR (ANother Tool for Language Recognition) to extract the Abstract Syntax Tree (AST) of code examples and then generates the Simplified Parse Tree (SPT) of the extracted structure. The system is query-based and relies on Minwise Hashing, which is based on four feature sets: Token, Parent, Sibling, and Variable Usages.

Zhang {\em et al.}~\cite{zhang2018efficient} introduced a recommendation system that uses both Locality Sensitive Hashing (LSH) and Collaborative Filtering (CF) simultaneously. To address issues related to time, space, and accuracy, they utilized a hybrid approach that combined minHash and SimHash to generate a signature matrix. The system then assigned signatures to buckets, grouping the most similar items in the same buckets.

Aytekin {\em et al.}~\cite{aytekin2019real} presented an updated version of Locality Sensitive Hashing (LSH) that can be used to develop a recommendation system with high accuracy, even when working with vast amounts of data. Their approach is faster than the standard LSH and recommends more diverse candidates.

\section{Methodology}

\begin{figure*}
    \centering
    \resizebox{12cm}{!}{
        \begin{tikzpicture}
          [node distance=1cm, start chain=going right,]
             \node[punktchain, join] (data) {Data Collection};
             \node[punktchain, join] (pre) {Data Pre-Processing};
             \node[punktchain, join] (extract) {Extraction of Code Examples};
             \node[punktchain][below right =1cm and -3.4151cm of data] (post){Model Assessment};
             \node[punktchain] (ml) {Recommending Code Examples Using LSH};
             \node[punktchain] (bert) {Applying Bert Model};
             \draw[->, thick] (extract.south) |-+(0,-1em)-| (bert.north);
             \draw[<-, thick] (bert.west) |-+(0,0)-| (ml.east);
             \draw[<-, thick] (ml.west) |-+(0,0)-| (post.east);
       \end{tikzpicture}
   }     
\label{fig:researchflow}
\caption{The main steps of the followed Methodology}
\end{figure*}
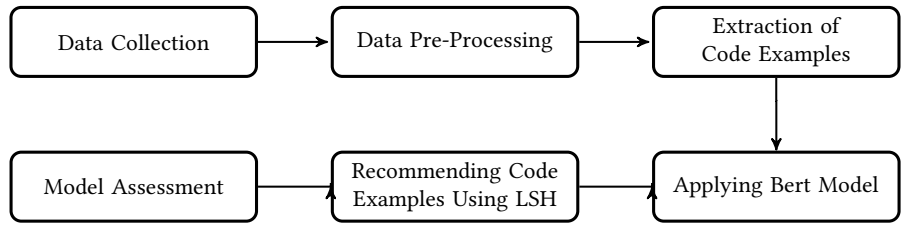

In this section, we will describe the main steps of our methodology for recommending code examples using Stack Overflow: Data collection, data preprocessing, Extraction of code examples, applying the BERT model, using LSH algorithms for recommending code examples and model assessment. 

\subsection{Data Collection}
Our input dataset consists of Stack Overflow posts, which are among the most widely-used sources for accessing discussions related to various API methods in different programming languages. To carry out our work, we downloaded Stack Overflow dump files. The dump files we obtained encompass all the discussions posted on Stack Overflow from 2008 to 2022. Similar to the approach followed in the study by Kim {\em et al.}~\cite{kim2018facoy}, we have retrieved the archived dump files of Stack Overflow posts and imported them into an SQL Server database.

\subsection{Data Pre-processing}
Once the dataset was imported into an SQL database for organizational purposes, we proceeded to clean the data. This involved pre-processing the data to prepare it for the next step. To achieve this, we followed a four-step process.

\textbf{Step 1:} When we attempted to filter Stack Overflow posts that contained Java code examples using the Java tag, we found that many posts labeled with Java also included code examples in other programming languages such as Javascript and C. Therefore, we filtered the posts to only include those that had Java code examples and excluded any that incorporated examples of other programming languages.

\textbf{Step 2:} Our database contains rows that represent posts and each row has 20 columns that display various aspects of a post. To obtain the posts necessary for our analysis, we have used the columns \emph{Id, AcceptedAnswerId,} and \emph{Score} for filtering. \emph{Id} is a post's unique identifier and \emph{AcceptedAnswerId} indicates the \emph{Id} of the post that is marked as the answer to a question post. Typically, most question posts have corresponding answer posts, and the \emph{AcceptedAnswerId} of the answer post is recorded in the database. The valid \emph{AcceptedAnswerId} values are stored in a table within the database for use in the subsequent steps.

\textbf{Step 3:} During this step, the answer posts are retrieved from the SQL database by filtering the posts where the "Id" equals the "AcceptedAnswerId". However, it is important to note that some questions on Stack Overflow may not have any answers, resulting in an empty "AcceptedAnswerId" field in those particular rows.

\textbf{Step 4:} After the filtration process in the previous steps, the resulting dataset contains Answer posts that have been specifically labeled by developers and their corresponding scores, indicating the level of approval by other users. we further applied a threshold on the score values and selected posts with scores greater than or equal to 2.

\subsection{Extraction of Code Examples from Informal Documentation}
Following the extraction of answer posts, it was necessary to extract code examples due to their integration with natural language comments. To achieve this, we filtered the code by identifying \verb!<pre><code>! and \verb!</code></pre>! tags within the posts to extract relevant examples. We replicated the approach suggested by Abdalkareem {\em et al.}~\cite{abdalkareem2017code} for code example extraction from Stack Overflow posts during this process. Table \ref{table:posts} demonstrates a summary of followed steps for filtering high-quality posts which were accomplished during steps 2 and 3 of our methodology.
\begin{table}
\caption {Data Extracted from Stack Overflow.}
\renewcommand{\arraystretch}{1.25}
\begin{center}
    \begin{tabular}{p{6.75cm} | c}
    \hline
    \textbf{Posts} & \textbf{Count} \\ \hline
    Posts with Java tags and not other programming languages & 377,517  \\
    Posts incorporating \verb!<code>! tag & 198,911 \\
	Code examples with more than 100 characters  & 128,688 \\
    Filtering non-code examples & 97,084 \\ 
    Code examples with scores greater than or equal to 2 & 61,361 \\
    \hline
    \end{tabular}
\end{center}
\label{table:posts}
\end{table}

\subsection{Applying BERT model}
We utilized the sentence\_transformers framework in Python to apply the \emph{BERT} model to our extracted code examples. This framework allows for the representation of text and images as dense numerical vectors. Our research used the pre-trained BERT model, which has been made publicly accessible in over 100 languages. We used this model to convert our code examples into numerical vectors. Each code example sample was fed as input to BERT, which combined token-based and semantic-based information to create a uniform (1*768) vector.
\subsection{Recommending Code Examples Using LSH}
After embedding code examples into numerical vectors, we have applied two \emph{LSH} algorithms, namely the {\em Random Hyperplane-based approach} and the {\em Query-aware} approach, to tackle our problem. The former assigns data samples into different buckets without taking the query into account, whereas the latter considers the query when assigning data samples. We incorporated both methods in our methodology and compared their outcomes, as previous research, has suggested that the Query-Aware approach (i.e., the second algorithm) performs better than Query-Oblivious approaches (i.e., the first algorithm) in retrieving data samples similar to a given query \cite{huang2015query}. In this paper, a query is a natural language sentence or a Java API method given by a developer as input, to find and recommend a list of relevant code examples.

\subsubsection{Random Hyperplane-based LSH}
To recommend code examples for API methods, we utilized Algorithm 1 from the work of Charikar {\em et al.}~\cite{charikar2002similarity} to perform dimensionality reduction of numerical vectors from high to lower dimensions. The resulting numbers were converted from decimal to binary values and sorted into buckets based on their binary values. This approach helped to find similar items to a given query, whether it be a natural language question or an API method, in a lower-dimensional space, using both data samples (code examples) and query vectors.

To explain the Random Hyperplane-based LSH algorithm, we followed the steps outlined in Algorithm 1. In this algorithm, data samples' vectors (D) are multiplied by randomized vectors (R), and the number of hash tables (M) determines how many times this process is repeated. This reduces the dimensionality of the data samples by mapping them to a lower dimension (k) (step 4). The resulting values are converted to binary values of 1 and 0 based on whether they are greater than or less than 0 (steps 5 and 6) and assigned to buckets based on their binary values (step 7).

The same process is applied to the query vector, and the resulting vector is also assigned to a bucket (steps 9 to 13). Data samples with the same bucket ID as the query vector are collected from all hash tables, and the cosine similarity between each data sample and the query vector is calculated and ranked based on their similarity values (from top to bottom) (step 15). Finally, the top N recommendations are returned to the user based on the number of recommendations (step 16).

To perform the mapping of data samples from a higher dimension to a lower one, we used random vectors of dimension (768 * d) that followed a Gaussian distribution. This is in line with Charikar's approach of using Gaussian distribution to create random vectors for mapping data samples to d-dimensional vectors.

\begin{algorithm}
\caption{\centering Random Hyperplane based LSH. Adapted from Charikar {\em et al.}~\cite{charikar2002similarity}}
\begin{algorithmic}[1]
\State For \emph{D} as Data samples, \emph{R} as a Randomized vector and \emph{M} as the number of hash tables;
\For{ $i \gets \texttt{1 to M}$}
    \State \textbf{For Data Samples:}
    \State $Result_{N*K}\gets D_{N*768} * R_{768*K}$;
    \State $SgnResult \gets Sign(Result_{N*K})$;
    \State if  the items of the SgnResult are negative assign 0, otherwise they are already 1;
    \State $BucketId \gets  \sum_{i=0}^{k-1} 2^i * SgnResult[i] $;
    \State \textbf{For Query Vector:}
    \State For Q as Query vector and R as a Randomized vector;
    \State $QResult_{N*K}\gets Q_{1*768} * R_{768*K}$;
    \State $SgnQRes \gets Sign(QResult_{N*K})$;
    \State if  the items of the SgnQRes are  negative assign 0, otherwise they are already 1;
    \State $QBucketId \gets  \sum_{i=0}^{k-1} 2^i * SgnQRes[i] $;
\EndFor
\State Retrieve the samples (u) that are at the same bucket as query vector (q) from all of the hash tables and rank them based on Cosine similarity; 
\State Retrieve top N similar samples as recommendation items;
\end{algorithmic}
\end{algorithm}
    


\subsubsection{Query-aware approach}
Query-oblivious methods assign data samples to buckets without taking into account the query, which may result in similar data samples being filtered out from the candidate set for the query. To address this issue, Huang {\em et al.}~\cite{huang2015query} proposed the Query-Aware LSH, where the query plays a role in assigning objects (data samples) to query bucket partitions. Similar to the Random Hyperplane-based algorithm, the Query-Aware algorithm is applied to code examples as data samples and natural language questions or API methods as queries. The Query-Aware approach resolves the problem of random shift that occurs when mapping data samples to a lower-dimensional space to assign them to buckets in traditional Query-Oblivious LSH algorithms. For example, E2LSH \cite{datar2004locality} applies a random shift (\emph{b}) after mapping the data sample into a lower-dimensional space. This random shift may negatively affect the process of finding the most similar data samples and assigning them to different buckets from the query. The Query-Aware approach simplifies the computation by eliminating the random shifting step \cite{huang2015query}.

Algorithm 2 represents data samples with \emph{D}, randomized vector with \emph{R}, number of hash tables with \emph{M}, and the threshold value with \emph{l}. The threshold value specifies the number of times that the Euclidean distance between a hash vector of a data sample and the query's hash vector is less than or equal to \emph{w/2}. In this algorithm, based on the number of hash tables, random vectors are generated. In each hash table, data samples are multiplied by their corresponding random vector to generate their hash values (step 3). Then, the Euclidean distance of each data sample's hash value is measured against the hash value of the query sample. If the distance value is less than or equal to \emph{w/2}, the occurrence number of these samples is incremented by 1 (step 4). If the occurrence number of a data sample (\ie{} code example) is greater than or equal to the threshold value \emph{l}, it is added to the candidate set (\emph{C}) (steps 5 to 7). After collecting the candidate samples, their Euclidean distance from the query vector is calculated and ranked incrementally (steps 9 to 10). Finally, the top N samples are recommended as the most similar code examples to the query (step 11). This approach filters the data samples based on the given query in two steps: (1) filtering based on the hash value similarity of the samples and the query and (2) selecting the filtered samples based on their similarity to the query.

\begin{algorithm}
\caption{\centering Query-Aware LSH. Adapted from Huang {\em et al.}~\cite{huang2015query}}
\begin{algorithmic}[1]
\State For \emph{D} as Data samples and \emph{R} as a Randomized vector, \emph{M} as the number of hash tables, \emph{l} as the threshold of the occurrence of a data sample ;
\For{ $i \gets \texttt{1 to M}$}
    \State Impose Hash functions by multiplying objects $(O_i)$ and query vector to Randomized vectors;
    \State Increase $ \#Col(O_i)$ if the $|H_i (O_i) - H_i (q)| <= w / 2 $;
    \If{$\#Col(O_i) \geq l$}
        \State $C = C \cup O_i$;
    \EndIf
    
\EndFor
\State Calculate the Euclidean distance between $O_i$ in C and q\; 
\State Sort the Euclidean distances incrementally; 
\State Retrieve the top N candidates as recommendation items from the sorted list;
\end{algorithmic}
\end{algorithm}


    


Both LSH-based algorithms were configured with specific parameter values, specifically \emph{K} = 10 and \emph{M} = 10. Notably, in the \emph{Query-Aware LSH} approach, the threshold values were equivalent to 0.1 of the maximum Euclidean distances between the query vectors and the data samples. These values were empirically determined due to their efficacy in achieving a harmonious trade-off between computational efficiency in processing times and the relevance level of suggested code examples.

\subsection{Model Assessment}
Once we implemented our LSH-based algorithms, which were the Random Hyperplane-based and Query-Aware approaches, we needed to evaluate their performance. To achieve this, and for comparison purposes, we utilized four different metrics: HitRate, Mean Reciprocal Rank (MRR), Average Execution time, and Relevance. HitRate and MRR were chosen based on their adoption by previous research \cite{zhou2019lancer}, while the Relevance metric was taken from PostFinder \cite{rubei2020postfinder}.

\section{Empirical Evaluation}
We have conducted two experiments to evaluate our LSH-based approach. Firstly, we compared two LSH-based algorithms, namely \emph{Random Hyperplane-based LSH} and \emph{Query-Aware LSH}, based on four metrics: \emph{Hit Rate}, \emph{Mean Reciprocal Rank}, \emph{Average Execution Time}, and \emph{Relevance}.

For the second part, we selected the algorithm that yielded better results for the aforementioned metrics and compared it with two state-of-the-art baselines, namely PostFinder and FaCoY, based on three additional metrics: \emph{Relevance}, \emph{Success Rate}, and \emph{Precision}.

\subsection{Variable Selection}
In this section, we will introduce and clarify the metrics that we utilized to evaluate our work. Our research focuses on the type of approach employed as the independent variable, with two distinct values for this factor:
\begin{itemize}
\item {\em Random Hyperplane-based LSH approach}
\item {\em Query-Aware LSH approach}
\end{itemize}

Within our study, we have examined several dependent variables, such as HitRate, MRR, Average execution time, and Relevance. In the following sections, we will provide a more detailed explanation of each of these metrics.\\
\textbf{\emph{HitRate}}: When given a set of queries (Q), the HitRate@k metric calculates the proportion of queries that have generated at least one relevant result among the top k recommended items. The following formula provides a definition for this parameter \cite{zhou2019lancer}:
\begin{equation}\label{E2LSH}
HitRate@K = \frac{1}{\lvert Q \lvert} \sum_{q\in Q} H(R(Q), k)  
\end{equation}
Based on the provided equation, Q represents a set of queries and \emph{H(R(Q), k)} is a function that returns 1 if at least one relevant item appears among the top k recommended items. Otherwise, it returns 0. The \emph{HitRate} metric is computed as the average of the 0 and 1 values, and a higher value of this metric indicates a more effective recommendation system. Specifically, if the \emph{HitRate} value is close to 1, it suggests that the recommendation system is successful.\\

\textbf{\emph{Mean Reciprocal Rank (MRR)}}: The \emph{MRR} metric is calculated as the average of the inverse of the first rank of the recommended items when the \emph{HitRate} occurs. To illustrate, suppose that two queries have been executed and the \emph{HitRate} of returned results for each of them is at the second and third ranks, respectively. In this case, the \emph{MRR} is the average of (1/2 + 1/3). The following formula defines the \emph{MRR} metric (\cite{zhou2019lancer}).
\begin{equation}\label{E2LSH-2}
MRR = \frac{1}{\lvert Q \lvert} \sum_{q\in Q} \frac{1}{\textit{First rank of the relevant result}}
\end{equation}

\textbf{\emph{Average Execution Time}}: Our approach's execution time is measured in two stages: the first stage involves measuring the time taken to apply hashing algorithms to data samples, and the second stage measures the time it takes to return results when hashing has already been applied. These measures are evaluated based on the number of hash tables utilized.\\
\textbf{\emph{Relevance}}: In our study, we use a metric called "Relevance" which indicates the score assigned by developers to the recommended items, specifically code examples in our case \cite{rubei2020postfinder}. The scores are determined according to  table \ref{tab:RelevanceSacle}.

\begin{table}
    \caption{The scoring scale of the Relevance metric}
    \label{tab:RelevanceSacle}
    \renewcommand{\arraystretch}{1.25}
    \begin{tabular}{c|p{7.25cm}}
    \hline
    {\bf Score} & {\bf Description}  \\\hline
        0 & No result has been returned \\
       1 & The result returned is not relevant \\
       2 & There are some hints but still out of context \\
       3 & The results incorporate some relevant results but not key features \\
       4 & The returned results are in the context of the query and are helpful \\
	  \hline
		\end{tabular}
\end{table}

\subsection{Study Design}
We used the randomized block design~\cite{basili1986experimentation} when conducting our survey with software developers. We divided the participants into three categories: junior, intermediate, and senior, based on their years of experience in software development. We generated a block by randomly selecting one person from each category. Then, the treatment options were randomly assigned within each block.

\subsection{Participants}
\begin{figure}
\centering
\includegraphics[width=0.9\columnwidth]{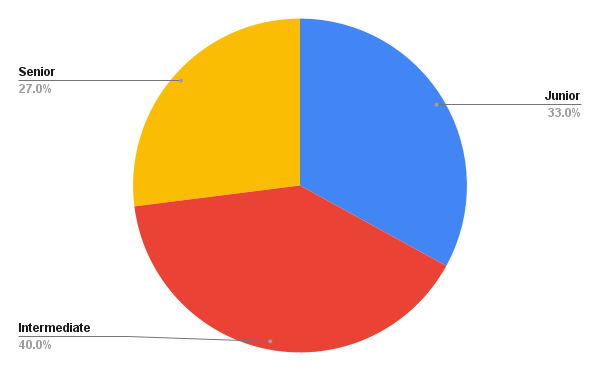}
\caption{Participants' skill levels. }
\label{fig:participants}
\end{figure}

The metrics were evaluated by a group of 15 software developers with varying levels of expertise, as shown in Figure \ref{fig:participants}, which depicts the participants ranging from junior to senior. Among the developers, 33\% had 1 to 2 years of Java work experience, 40\% had 2 to 3 years of experience, and 27\% had over 3 years of experience. The metrics' values were determined based on the highest percentage values.
For example, if the Relevance metric is considered, and 5\% of the participants chose a score of 0, 5\% selected a score of 1, 20\% chose a score of 2, 45\% selected a score of 3, and 25\% chose a score of 4 (as shown in Table \ref{tab:RelevanceSacle}), a Relevance value of 3 would be recorded. During the experiment, the developers evaluated the recommendation items individually based on the above-mentioned metrics.
\subsection{Analysis Method}
To compare metrics between two algorithms with numerical values, such as relevance, we used the Wilcoxon Rank Sum Test \cite{marascuilo1988statistical, siegel1956nonparametric, wilcoxon1992individual}. This non-parametric test was chosen because our data did not follow a normal distribution. The test works by ranking the data samples from lowest to highest and summing the ranks of the two groups. It evaluates whether there is a significant difference between the medians of the two groups.\\
Next, we examined the hypotheses by comparing the obtained values with predetermined threshold values \cite{wilcoxon1992individual}. Two hypotheses were formulated for the comparison, with the first one assuming no difference between the two dependent groups, and the alternative hypothesis assuming a difference between the two dependent groups. We utilized the Wilcoxon Rank Sum test to evaluate the Relevance metric for both algorithms investigated in our list of queries, which consisted of either natural language queries or Java API methods. A detailed discussion of this will be presented in the Results section.

\subsection{Comparison of our work with PostFinder and FaCoY}
In order to compare our approach with baselines, we need to measure the defined metrics in PostFinder\cite{rubei2020postfinder}. The main difference between our work and these studies is that they measure the introduced metrics based on the recommended Stack Overflow posts, which include both comments and code snippets. However, our work measures those metrics based solely on code snippets. The primary reason for selecting PostFinder and FaCoY as our baselines is their common data source, Stack Overflow. Additionally, we used 50 queries that PostFinder utilized for evaluation in configuration G to compare itself with FaCoY. These 50 queries were selected from the 10 most popular Java libraries: Jackson, SWT, MongoDB driver, Javax Servlet, JDBC API, JDT core, Apache Camel, Apache Wicket, Twitter4j, and Apache POI \cite{rubei2020postfinder}.

To further strengthen the validity of our comparison, it is important to acknowledge that PostFinder utilized the dump files of Stack Overflow up until June 2017. In order to maintain consistency and ensure an accurate evaluation, we have specifically filtered the code examples for this part, considering only those with dates up until June 2017. Below are the comparison metrics:

\textbf{\emph{Relevance}}: This is the score given to the the pair of \textless query, recommended code examples\textgreater \space based on the table \ref{tab:RelevanceSacle}.

\textbf{\emph{Success Rate}}: The query is considered relevant if at least one of the top 5 retrieved code examples has a score of 3 or 4. The success rate is determined by dividing the number of relevant queries by the total number of queries.

\textbf{\emph{Precision}}: The metric calculates the ratio of pairs that receive a score of 3 or 4 in the top 5 recommendations to the total number of pairs.\\
According to Table \ref{tab:CompWithBaselines}, the results of these 3 measures for PostFinder and FaCoY are extracted from \cite{rubei2020postfinder}. 


\section{Results}
There are two types of queries \ie{} Natural Language and API Names-based queries that we have considered for the comparison of LSH-based algorithms that are listed in Table \ref{tab:Queries}. We selected these query variants to examine the extent to which the BERT model could find more relevant code examples when the query is presented in natural language or a Java API format. These are the queries that are extracted from other publications \cite{kim2018facoy,diamantopoulos2018codecatch}. \\
\begin{table*}
\caption {Different query types.}
\label{tab:Queries}
\renewcommand{\arraystretch}{1.25}
\begin{center}
    \begin{tabular}{l | l | l }
    \hline
    \centering{\textbf{Query ID}} & {\textbf{Natural Language based}} & {\textbf{API Names based}} \\ \hline
    Query 1 & How to add an image to a JPanel?  & Jpanel.add()\\
 Query 2 & How to generate a random alpha-numeric string? & StringBuilder.append()\\
 Query 3 & How do I create a file and write to it? & Writer.write() \\
 Query 4 & How do I invoke a Java method when given the method name as a string? & Method.invoke()\\
 Query 5 & Remove HTML tags from a String? & Jsoup.parse()\\
 Query 6 & How to get the path of a running JAR file? & URLDecoder.decode()\\
 Query 7 & Getting a File’s MD5 Checksum in Java & MessageDigest.digest()\\
 Query 8 & Loading a properties file from Java package & Properties.load()\\
 Query 9 & How can I play sound in Java?& AudioSystem.getAudioInputStream()\\
 Query 10 & What is the best way to SFTP a file from a server? & JSch.getSession()\\
 Query 11 & How to read a CSV file? & BufferedReader.readLine()\\
 Query 12 & How to generate MD5 hash code? & MessageDigest.getInstance()\\
 Query 13 & How to send a packet via UDP? & DatagramSocket.send()\\
 Query 14 & How to split a string? & String.split()\\
 Query 15 & How to play an audio file? & MediaPlayer.play()\\
 Query 16 & How to upload a file to FTP?& FTPClient.storeFile()\\
 Query 17 & How to initialize a thread? & thread.start()\\
 Query 18 & How to connect to a JDBC database? & DriverManager .getConnection()\\
 Query 19 & How to read a ZIP archive? & zipFile.getInputStream()\\
 Query 20 & How to send an email? & MimeMessage.setFrom()\\

    \hline
    \end{tabular}
\end{center}
\label{table:questions}
\end{table*}

\textbf{RQ1:} What is the performance comparison between the Random Hyperplane-based LSH and the Query-Aware LSH approaches in recommending code examples, considering metrics such as Hit Rate, Mean Reciprocal Rank, Average execution time, and Relevance?\\

Table \ref{table:metrics} demonstrates the values of HitRate (3rd to 5th columns), MRR (6th column), and Relevance (7th column) metrics. According to this table, for natural language-based queries, the \emph{HitRate} values are (0.2, 0.5, 0.55) and (0.5, 0.8, 0.9) respectively for the top 10, 20, and 30 recommendations generated by \emph{Random Hyperplane-based LSH} and \emph{Query-Aware LSH}. The results indicate that \emph{Query-Aware LSH} gives better results than \emph{Random Hyperplane-based LSH} in terms of \emph{HitRate}. This is because \emph{Query-Aware LSH} is capable of identifying more relevant code examples based on the given query. As the number of recommended examples increases, the likelihood of finding relevant code also increases, resulting in a higher \emph{HitRate} value.

Additionally, Table \ref{table:metrics} presents a notable discovery regarding the \emph{MRR} measure, which reveals a difference in performance depending on the type of query used. Specifically, both the \emph{Random Hyperplane-based LSH} and \emph{Query-Aware LSH} algorithms perform better in terms of \emph{MRR} values when handling API method names as queries, compared to Natural Language-based queries. The two algorithms demonstrate almost identical \emph{MRR} values for the same query type. A higher \emph{MRR} value implies that the algorithm returns relevant code examples at a lower rank, while a lower \emph{MRR} value indicates that the algorithm returns relevant code examples at higher ranks or fails to return any relevant results.

Moreover, Table \ref{table:metrics} provides a summary of the \emph{Relevance} metric results for the various types of queries and algorithms evaluated. According to our findings, natural language-based queries produce higher \emph{Relevance} scores when recommending code examples. This result can be attributed to the use of the \emph{BERT} model, which considers both the semantic and contextual information of queries and code examples \cite{devlin2018bert}. As a result, it performs well when processing natural language text compared with API Names.\\
\begin{table*}
\caption {Summary of Results of HitRate, MRR and Relevance metrics for different query types.}
\renewcommand{\arraystretch}{1.25}
\begin{center}
\begin{tabular}{ p{2.75cm}|p{2.75cm}|c|c|c|c|c }
 \hline
 \textbf{Query Type} &\textbf{Approaches} & \textbf{HitRate (Top 10)} & \textbf{HitRate (Top 20)} & \textbf{HitRate (Top 30)} & \textbf{MRR} & \textbf{Relevance} \\
 \hline
 \emph{Natural Language-based Queries} & \emph{Random Hyperplane-based LSH}
   & 0.2    &0.5&   0.55 & 0.0774 & 1.7\\
  & \emph{Query-Aware LSH}    & 0.5    &0.8&   0.9 &0.2582 & 2.7\\
  \hline
   \emph{API Names-based Queries} & \emph{Random Hyperplane-based LSH}   & 0.15    &0.25&   0.3  & 0.07142 & 2.05\\
  & \emph{Query-Aware LSH}    & 0.5    &0.5&   0.5 & 0.246 & 2.65\\
 \hline
\end{tabular}
\end{center}
\label{table:metrics}
\end{table*}
Table \ref{tab:AvgTime} presents the average execution times for the LSH creation and recommendation processes of both the \emph{Random Hyperplane-based LSH} and the \emph{Query-Aware LSH} algorithms.
We conducted measurements on these elements using varying numbers of hash tables, ranging from 2 to 50. The number of hash tables determines how frequently the hashing algorithm is applied to both the query and data samples. Increasing the number of hash tables leads to more data points being mapped to the same bucket as the query, resulting in a higher likelihood of finding more similar data samples and therefore improving the accuracy of the nearest neighbor search. However, a larger number of hash tables can also increase the computational complexity. Furthermore, Table \ref{tab:AvgTime} displays the average execution time for all considered queries, revealing that the \emph{Query-Aware LSH} algorithm outperforms the \emph{Random Hyperplane-based} algorithm in both the LSH creation phase and the recommendations generation phase. On average, the \emph{Query-Aware LSH} algorithm took 11.21 seconds to create hash tables and less than 1 millisecond to generate code examples recommendations. Conversely, the \emph{Random Hyperplane-based} algorithm took 49.56 seconds on average to create hash tables and 0.98 seconds on average to generate code examples recommendations.

\begin{figure*}
    \centering
    \subfigure[Creation time based on the number of Hash Tables.]{\includegraphics[width=0.5\textwidth]{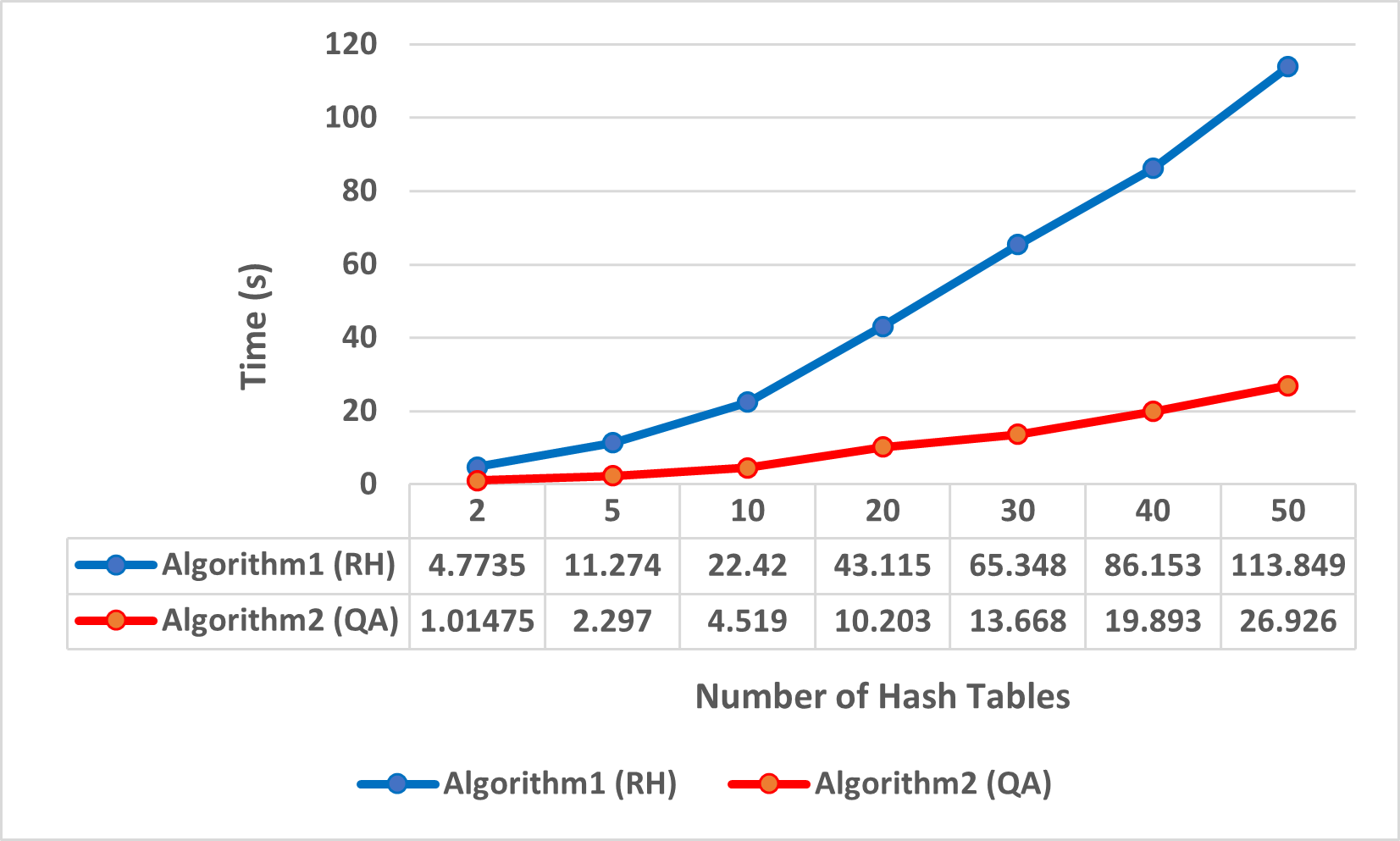}}\hfill
    \subfigure[Boxplot of Creation time based on the number of Hash Tables.]{\includegraphics[width=0.5\textwidth]{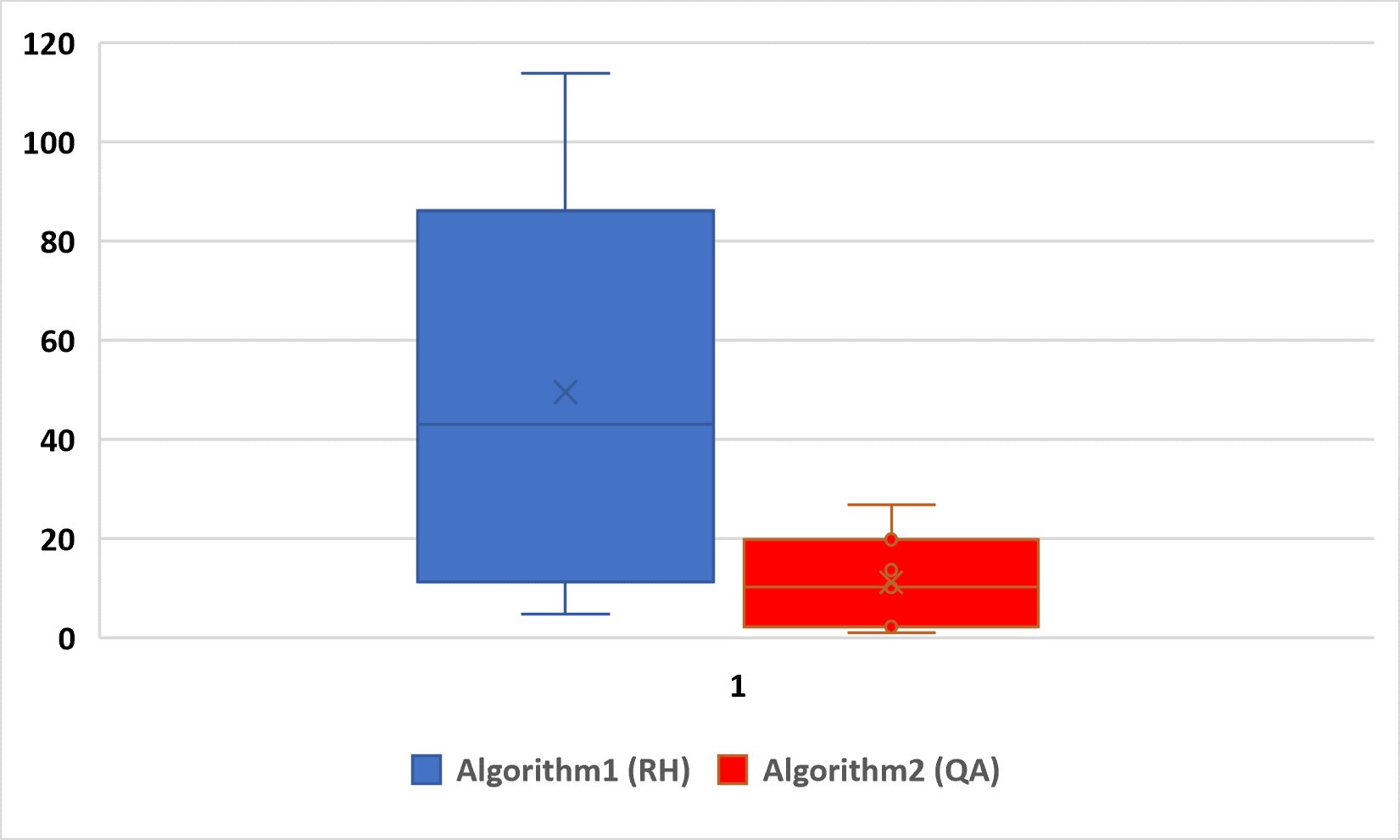}}\hfill
    \subfigure[Recommendation time based on the number of Hash Tables.]{\includegraphics[width=0.5\textwidth]{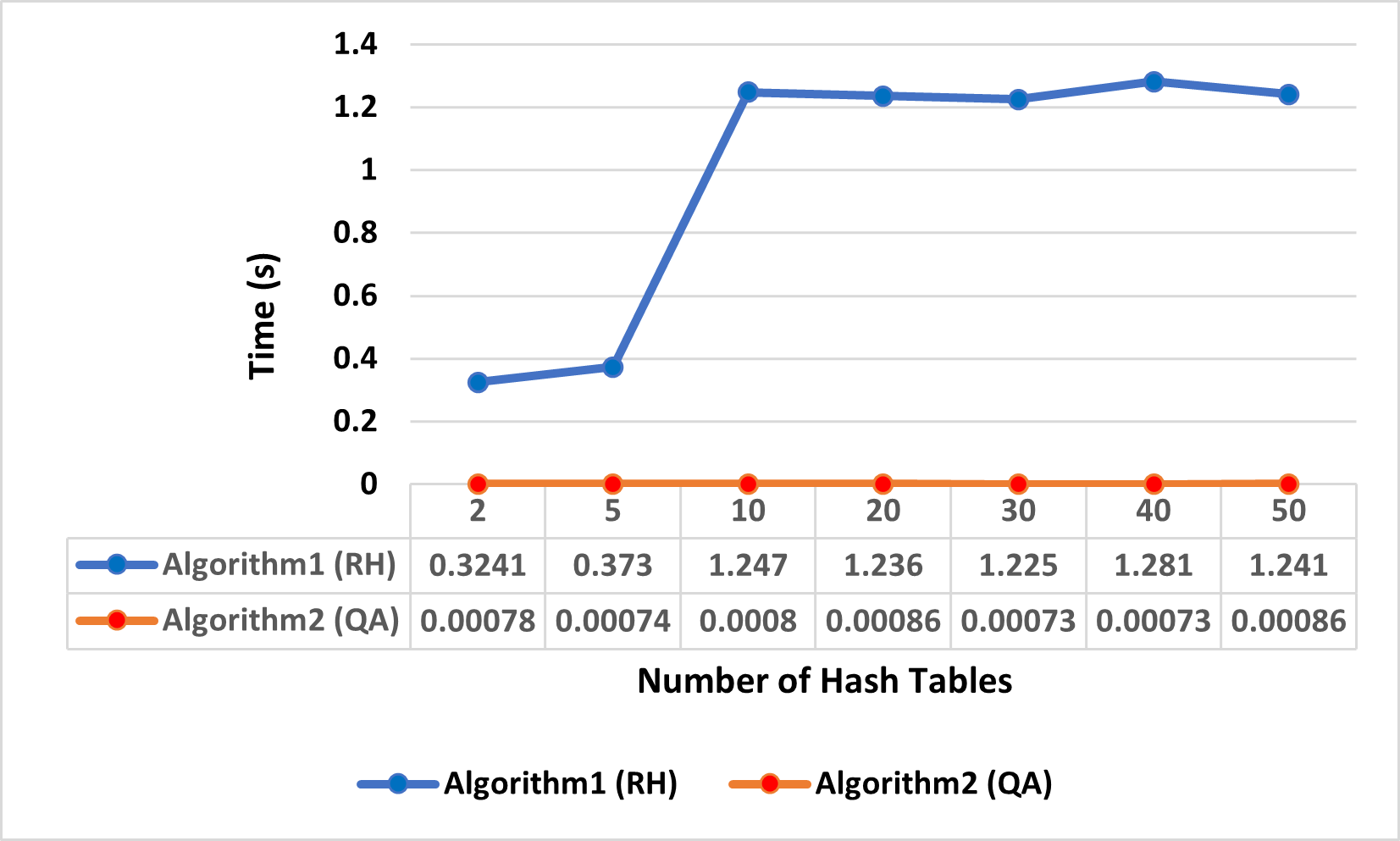}}\hfill
    \subfigure[Boxplot of Recommendation time based on the number of Hash Tables.]{\includegraphics[width=0.5\textwidth]{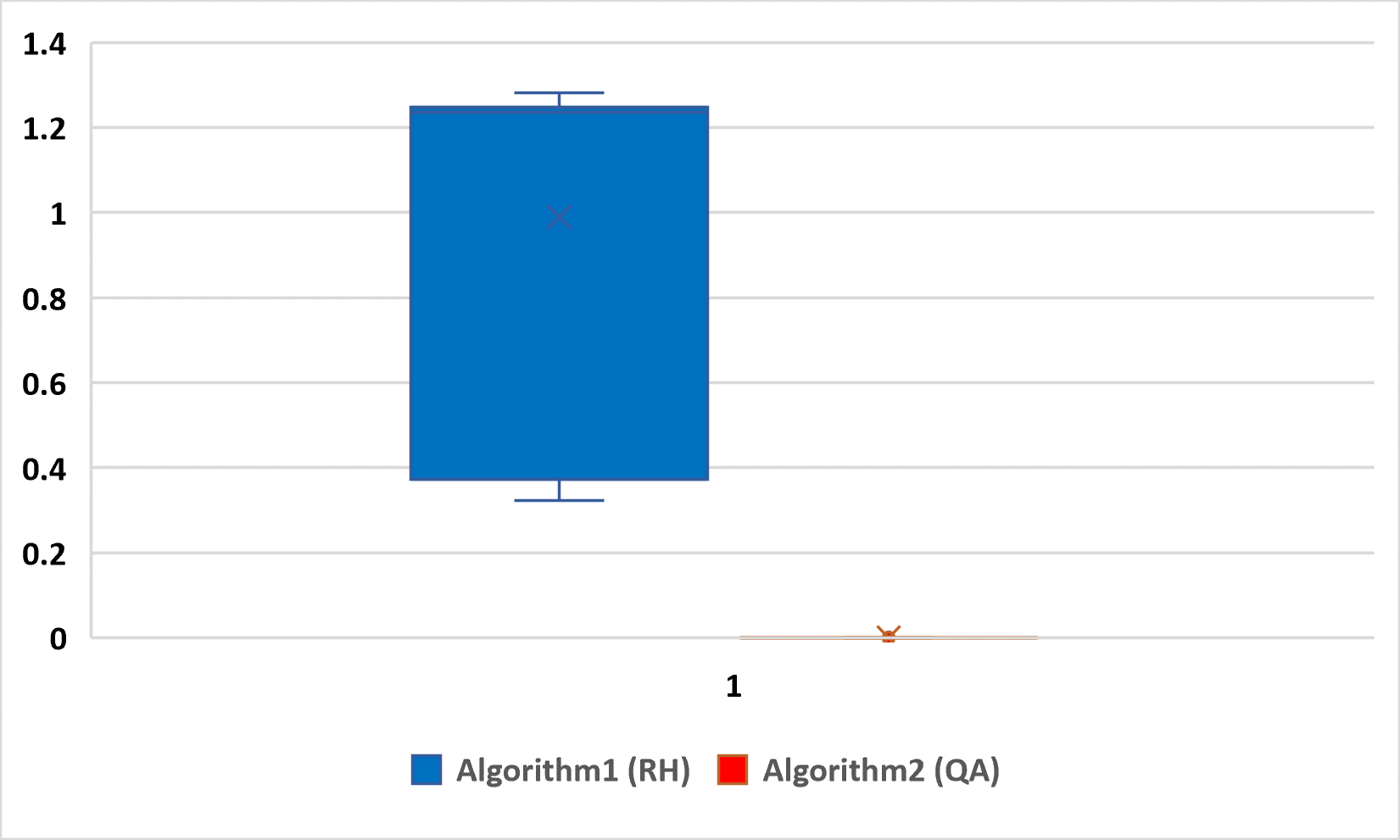}}\hfill
    \caption{LSH figures adapted from Rahmani \cite{rahmani2023towards}.}
    \label{fig:LSH}
\end{figure*}

\begin{table*}
\caption {The Average Execution Time of \emph{Random Hyperplane-based LSH} (1) and \emph{Query-Aware LSH} (2) in seconds.}\label{tab:AvgTime}
\renewcommand{\arraystretch}{1.25}
\begin{center}
\begin{tabular}{l|l|l|l|l}
		\hline
			{\bf Number of hash tables} & {\bf LSH creation time(1)}  & {\bf Recommendation time(1)} & {\bf LSH creation time(2)}  & {\bf Recommendation time(2)} \\
	  \hline
			2 & 4.7735 & 0.3241 & 1.01475 & 0.000779 \\
			5 & 11.274 & 0.373 & 2.297 & 0.0007476 \\
			10 & 22.420 & 1.247 & 4.519 & 0.0008069 \\
			20 & 43.115 & 1.236 & 10.203 & 0.0008623 \\
			30 & 65.348 & 1.225 & 13.668 & 0.0007353 \\
			40 & 86.153 & 1.2814 & 19.893 & 0.0007533 \\
			50 & 113.849 & 1.241 & 26.926 & 0.000861\\
	  \hline
		\end{tabular}
\end{center}
\label{table:questions-2}
\end{table*}

Figure \ref{fig:LSH} (a) illustrates the time of the creation of Hash tables, along with the allocation of data samples to their respective buckets, for the two algorithms proposed, namely the \emph{Random Hyperplane-based} and the \emph{Query-Aware LSH}. It can be observed from the figure that the range of values for this metric for these algorithms are (4.77, 113.85) and (1.014, 26.926) respectively. Figure \ref{fig:LSH} (b) displays the Boxplot representation of the LSH creation time for the two algorithms being considered in this study, namely, the \emph{Random Hyperplane-based} and the \emph{Query-Aware LSH} algorithms. The figure shows that the first quartile, median, and third quartile values for the first and second algorithms are (11.274, 43.115, 86.153) and (2.297, 10.203, 19.893), respectively. Moreover, the range between the Maximum and Minimum values for the two algorithms are 109.0755 and 25.912, respectively.

The recommendation time for the \emph{Random Hyperplane-based} algorithm fluctuates between 0.3241 seconds to 1.241 seconds as the number of Hash tables increases from 2 to 50, as shown in Figure \ref{fig:LSH} (c). On the other hand, the recommendation time for the \emph{Query-Aware LSH} algorithm ranges from 0.00077 seconds to 0.000861 seconds, which is significantly lower than that of the \emph{Random Hyperplane-based} algorithm. Figure \ref{fig:LSH} (d) illustrates the Boxplot of the recommendation time for both the \emph{Random Hyperplane-based} and \emph{Query-Aware LSH} algorithms. As per the plot, the first Quartile, Median, and third Quartile values for the \emph{Random Hyperplane-based} algorithm are (0.373, 1.236, 1.2814), while for the \emph{Query-Aware LSH} algorithm, the values are (0.0007476, 0.0008623, 0.0007533). Additionally, the difference between the Maximum and Minimum values for the two algorithms is 0.9573 and 0.0001257, respectively.\\
The results discussed can be accounted for by the approach of Query-Aware LSH, which doesn't necessitate placing all data samples into buckets. Instead, it focuses on only those samples whose hash values are in proximity to the query's hash value. Additionally, the Query-Aware LSH algorithm decreases the search space by excluding data samples whose hash vectors are near the query during the LSH creation. This takes place at the initial stage of the search process, resulting in the algorithm's ability to identify relevant samples more effectively \cite{huang2015query}.

\textbf{RQ2:} Does the Random Hyperplane-based LSH approach differ from the Query-Aware LSH approach in terms of Relevance when recommending code examples?\\

The second research question was addressed by formulating a null hypothesis and an alternative hypothesis, as stated below:
\begin{itemize}
    \item {$H_{0{2}}$}: There is no statistically significant difference between the  \emph{Query-Aware LSH} and 
 the \emph{Random Hyperplane-based LSH} in terms of \emph{Relevance}. 
    \item {$H_{a{2}}$}: There is a statistically significant difference between the \emph{Query-Aware LSH} and the \emph{Random Hyperplane-based LSH} in terms of  \emph{Relevance}.
\end{itemize}

\begin{table*}
\caption {The result of a Wilcoxon Rank Sum test based on the Relevance metric for different query types} \label{tab:Wilcoxon}
\renewcommand{\arraystretch}{1.25}
\begin{center}
\begin{tabular}{ l|c|c|c|c|c|c}
 \hline
 \textbf{Query Type} & \textbf{Critical U-value} &  \textbf{U-value} & \textbf{Z-Score} & \textbf{Critical P-value} & \textbf{P-value} & \textbf{Result}   \\
 \hline
 \emph{Natural Language-based} &  127  & 107 & -2.5 & 0.05 & 0.012 & Significant\\
   \emph{API Names-based} &   37  & 47 & -1.41 & 0.05 & 0.16 & Not Significant   \\
 \hline
\end{tabular}
\end{center}
\label{table:questions-3}
\end{table*}

\begin{table*}
\caption {Comparison of Query Aware LSH (QALSH) with PostFinder and FaCoY based on three metrics} \label{tab:CompWithBaselines}
  \centering
  \renewcommand{\arraystretch}{1.2}
  \begin{tabular}{p{2cm}|c|c|c|c|c|c|c|c|c}
    \hline
    \multirow{2}{*}{\textbf{}} & \multicolumn{3}{c|}{\textbf{Success rate}} & \multicolumn{3}{c|}{\textbf{Precision}} & \multicolumn{3}{c}{\textbf{Relevance}}   \\
    \cline{2-10}
    & \textbf{QALSH} & \textbf{PostFinder} & \textbf{FaCoY} & \textbf{QALSH} & \textbf{PostFinder} & \textbf{FaCoY} & \textbf{QALSH} & \textbf{PostFinder} & \textbf{FaCoY} \\
    \hline
    Mean & 0.83 & 0.95 & 0.77 & 0.51 & 0.66 & 0.33 & 2.64 & 2.78 & 2.09   \\ \hline
    Std & 0.38 & 0.22 & 0.42 & 0.25 & 0.30 & 0.26 & 1.06 & 1.01 & 1.01 \\ \hline
    First quartile & 1.0 & 1.0 & 1.0 & 0.2 & 0.40 & 0.20 & 2.0 & 2.0 & 1.0  \\ \hline
    Second quartile & 1.0 & 1.0 & 1.0 & 0.4 & 0.60 & 0.40 & 3.0 & 3.0 & 2.0  \\ \hline
    Third quartile & 1.0 & 1.0 & 1.0 & 0.8 & 1.0 & 0.45 & 4.0 & 4.0 & 3.0  \\
    
    \hline
  \end{tabular}
\end{table*}


We present the findings from our evaluation of two algorithms: the \emph{Random Hyperplane-based LSH} and the \emph{Query-Aware LSH}. We conducted a comparison between these two algorithms using the Wilcoxon Rank Sum Test. Specifically, we examined whether there was a significant difference between the algorithms in terms of their ability to provide relevant code examples for both natural language-based and API method names-based queries.

Table \ref{tab:Wilcoxon} presents a summary of the "Relevance" outcomes for queries based on natural language and API methods' names. For natural language-based queries, the U-value is 107, the U-value is lower than the critical U-value, which indicates that the results are not due to chance and there is a significant difference between the two algorithms in terms of Relevance. Moreover, as shown in table \ref{tab:Wilcoxon} for the API methods' names-based queries, since the U-value is higher than the critical U-value, the differences between the two groups are more likely to be attributed to chance. Therefore, the null hypothesis cannot be rejected in this case.\\

\textit{\textbf{RQ3:}} How does the proposed algorithm perform compared to the recent state-of-the-art methods, namely PostFinder and FaCoY?

Since the \emph{Query-Aware LSH} (QALSH) yielded higher values for the evaluated metrics compared to the \emph{Random Hyperplane-based LSH}, we selected this algorithm for comparison with the baselines, namely Postfinder and FaCoY. According to Table \ref{tab:CompWithBaselines}, the average values of success rate, precision, and relevance for QALSH are 0.83, 0.51, 2.64, while the corresponding values for FaCoY are 0.77, 0.33, 2.09 respectively. PostFinder achieved higher values for these metrics, which are 0.95, 0.66, 2.78. Hence, QALSH outperformed FaCoY in three metrics. However, PostFinder achieved better results than QALSH in three measures, even though their average relevance values were closely matched 2.64 vs 2.78.

\section{Threats to validity}
Despite the careful execution of our evaluation step, there are certain factors that could potentially compromise the validity of our empirical evaluation. These threats to validity can be categorized into three distinct categories \cite{Godwin}.

\subsection{Internal Validity}
Potential threats to internal validity in our study include the pre-processing of data samples, which involved filtering based on regular expressions, code example length, and post-score. This process may have inadvertently excluded valid API methods used in some posts with short code examples (one or two lines). As a result, the accuracy of our outcomes could be compromised and necessitates careful consideration. Additionally, code examples demonstrating correct API method usage may have been filtered out due to low scores below the specified threshold values, further posing a threat to result validity.

\subsection{External validity}
Although we tested our algorithms on Java code examples sourced from Stack Overflow posts, it may be necessary to assess their performance on posts containing code examples in different programming languages, such as C/C++ or Python. Doing so would enable us to gauge the ability of our LSH-based algorithms to retrieve relevant code examples in various programming languages and assess their performance based on predefined metrics.

\balance
\subsection{Conclusion Validity}
We provided all the necessary details for replicating our study in the online appendix\footnote{https://github.com/icpc2024-so/ICPC-2024}. However, applying our results to other domains requires careful analysis of that domain to address any uncertainties.

\section{Conclusion and future work}
Based on our findings, it appears that utilizing natural language-based queries yields more accurate search results compared to API Names-based queries. Additionally, the recommendation system proves to be more effective in retrieving relevant information when using natural language queries. When analyzing both types of queries, the Query-Aware LSH consistently outperforms the Random Hyperplane-based LSH in terms of metrics such as HitRate, MRR (Mean Reciprocal Rank), Average Execution time, and Relevance. Through the use of the Wilcoxon Rank Sum test, we determined that the Query-Aware LSH performs better than the Random Hyperplane-based LSH specifically when the query is in a natural language format, but not for API Names-based queries. Moreover, the comparison of QALSH with PostFinder and FaCoY demonstrates its capability to recommend valuable code examples sourced from Stack Overflow.

In the future, we aim to advance our approach and transform it into a recommendation system integrated as a plugin in the Eclipse IDE. This integration will enable the system to assist developers with their ongoing tasks within the IDE. We also plan to conduct large-scale controlled experiments, involving software developers utilizing our recommender system for their software engineering and evolution tasks. Additionally, we intend to explore informal documentation sources like email discussions, forums, and bug reports to gather more code examples for enhanced recommendations.\\




\end{document}